\documentclass[journal,draftcls,12pt,onecolumn,oneside]{IEEEtran}
\usepackage{mathrsfs}
\usepackage{pifont}
\usepackage{bbding}
\usepackage{amsfonts}
\usepackage{amssymb}
\usepackage{amsmath}
\usepackage{graphicx}
\usepackage{subfigure}
\usepackage{cite}
\usepackage{bbding}
\usepackage{multirow}
\usepackage{mathtools}
\usepackage{array}
\usepackage{epstopdf}
\usepackage{color}
\begin{document}

\title{Energy-Efficient Resource Allocation in NOMA Heterogeneous Networks}

\author{Haijun Zhang,~\IEEEmembership{Senior Member,~IEEE}, Fang Fang,~\IEEEmembership{Student Member,~IEEE}, Julian Cheng,~\IEEEmembership{Senior Member,~IEEE}, Keping Long,~\IEEEmembership{Senior Member,~IEEE},
Wei Wang,~\IEEEmembership{Senior Member,~IEEE},  and Victor C.M. Leung,~\IEEEmembership{Fellow,~IEEE}
\thanks{Haijun Zhang and Keping Long are with the Engineering and Technology Research Center for Convergence Networks, University of Science and Technology Beijing, Beijing, 100083, China (e-mail: haijunzhang@ieee.org, longkeping@ustb.edu.cn).

Fang Fang and Julian Cheng are with School of Engineering, The University of British Columbia, Kelowna, BC, V1X 1V7, Canada (e-mail: fang.fang@alumni.ubc.ca, julian.cheng@ubc.ca).

Wei Wang is with the College of Information Science and Electronic Engineering, Zhejiang University, Hangzhou 310027, China, and also with the Zhejiang Provincial Key Laboratory of Information Processing, Communication and Networking, Hangzhou 310027, China (e-mail: wangw@zju.edu.cn).

Victor C.M. Leung is with the Department of Electrical and Computer Engineering, The University of British Columbia, Vancouver, BC V6T 1Z4 Canada (e-mail: vleung@ece.ubc.ca).

}} \maketitle

\begin{abstract}
Non-orthogonal multiple access (NOMA) has attracted much recent attention owing to its capability for improving the system spectral efficiency in wireless communications. Deploying NOMA in heterogeneous network can satisfy users' explosive data traffic requirements, and NOMA will likely play an important role in the fifth-generation (5G) mobile communication networks.  However, NOMA brings new technical challenges on resource allocation due to the mutual cross-tier interference in heterogeneous networks. In this article, to study the tradeoff between data rate performance and energy consumption in NOMA, we examine the problem of energy-efficient user scheduling and power optimization in 5G NOMA heterogeneous networks. The energy-efficient user scheduling and power allocation schemes are introduced for the downlink 5G NOMA heterogeneous network for perfect and imperfect channel state information (CSI) respectively. Simulation results show that the resource allocation schemes can significantly increase the energy efficiency of 5G NOMA heterogeneous network for both cases of perfect CSI and imperfect CSI.

\end{abstract}

\section{Introduction}

In order to meet the explosive growth of wireless traffic data requirement and overcome the shortage of the frequency resource, non-orthogonal multiple access (NOMA) has been proposed as a promising technique in the fifth generation (5G) of wireless communication systems \cite{LinglongDaiNOMA2015}. Different from orthogonal frequency division multiple access (OFDMA) technique using orthogonal subcarriers, NOMA enables multiple users to be multiplexed on the same frequency band by applying the successive interference cancelation (SIC) technique at the receivers \cite{DingIEEEMag16}. According to the NOMA protocol, the user who has worse channel state information (CSI) is assigned more power than the user who has better CSI \cite{LinglongDaiNOMA2016,Chen2017}. This protocol guarantees the weak users' access by successively decoding the signals from the received signals in the decreasing order of power. It was shown that the NOMA scheme can achieve higher spectral efficiency and higher data rate than the traditional OFDMA scheme \cite{ZhiguoDingNOMA2016}.

Driven by the rapid increase of wireless terminal equipments and wide usage of mobile Internet, heterogeneous networks have
emerged as one of the most promising network infrastructures to provide high system throughput and large coverage of indoor
and cell edge scenarios in 5G wireless communication systems \cite{HaijIEEEHet15}. In such an architecture, a macrocell is
overlaid by several small cells, e.g., microcell, picocell and femtocell, to significantly improve the system throughput
and the spectral efficiency. For heterogeneous networks, frequency band sharing between macrocell and small cells is viable,
and it is also more efficient to reuse the frequency bands within a macrocell \cite{HaijIEEEsmallcell15}.
However, the cross-tier interference can severely degrade the quality of the wireless transmission. The advantage of the heterogeneous network also comes with some fundamental challenges on cross-tier interference mitigation, which have been studied in previous research by some techniques, e.g. precoding technique and resource management \cite{HaijIEEEsmallcell15}. In this article, we apply NOMA with SIC technology to heterogeneous networks in order to alleviate the cross-tier interference and improve the system energy efficiency via resource optimization. In the traditional OFDMA heterogeneous networks, the frequency band can be divided into several sub frequency bands, and users in the macro cell and small cells are assigned to different sub frequency bands in order to avoid the cross-tier interference \cite{HaijIEEEsmallcell15}. However, in NOMA heterogeneous networks, SIC is applied at the receivers to allow multiple users to be multiplexed on the same sub frequency band and the strong users with higher CSI can cancel the interference from the weak user with lower CSI multiplexed on the same sub frequency band.

With the overwhelming increase of the traffic data and mobile devices, the energy cost has rapidly increased and become an important issue in the green cellular network because of the increasing amount of $\rm CO_2$ emission levels caused by energy consumption \cite{globalIEEEComMag11,YChenIEEComMag11,GAueIEEEWComMag11,ZHasanIEEEComSurvey11}. Thus, fast growing energy consumption and limited global energy resources are the important motivation for the research on energy efficient communications in NOMA heterogeneous networks.

Different from the existing works in NOMA networks, in this article, we focus on the energy efficient wireless communications for the downlink 5G NOMA heterogeneous networks via resource optimization. More specifically, we study user scheduling and power allocation by considering both perfect CSI and imperfect CSI. We first present network architecture of a 5G NOMA heterogeneous network. Then we describe an effective resource allocation scheme in 5G NOMA heterogeneous networks to improve the system energy efficiency with perfect CSI. A user scheduling and power optimization scheme with imperfect CSI is also introduced. Finally, we conclude this article.

\section{Resource Optimization for NOMA Heterogeneous Networks}

In NOMA heterogeneous networks, a macrocell with one macro base station (BS) is overlaid by several small cells where each small cell implements NOMA technique. In this system, macrocell users (MUEs) and small cell users (SUEs) can be served on the same sub frequency band at the same time to achieve high spectral efficiency. In other words, the overlay frequency band sharing model implies that macrocell users can access the sub frequency band freely and efficiently under the minimum data rate requirement determined by the quality of service (QoS). High data rate can be achieved by allocating more power to the UEs; however, this will incur high power consumption. Hence, finding the tradeoff between the data rate improvement and power consumption has become an important and inevitable trend in the next generation of wireless communication systems. For energy efficient wireless communications, we aim to design a resource scheduling scheme to balance the system sum rate improvement and the power consumption.
\subsection{NOMA Overview}
We now present an overview of the resource allocation for NOMA systems. Different from the conventional OFDMA technique, multiple users in NOMA system can share the same frequency band instead of using orthogonal spectrum. By applying SIC technique at the receivers, a user in NOMA system can successively extract the intended signal from the received aggregated signals. In \cite{IEEESystem13}, SIC was exploited to decode the signals from the received signals in the decreasing order of power. Therefore, the user who has higher channel gain can cancel the interference signal from the user who has lower channel gain. NOMA can be combined with MIMO to further improve capacity \cite{DingMIMOIEEE16}. Particularly, the pioneering work of beamspace MIMO-NOMA was firstly proposed to integrate NOMA for millimeter wave massive MIMO systems in \cite{LinglongDaiNOMA2016}, where a dynamic power allocation scheme with low complexity was proposed by solving the challenging joint power optimization problem. The statistical channel information was applied at the transmitter to achieve the maximum of  system energy efficiency. The energy-efficient resource allocation for a downlink NOMA single cell network has been investigated in \cite{FangIEEETrans16}. The authors in \cite{FangIEEETrans16} applied matching theory to assign the users to the different subchannels, and utilized difference of convex functions (DC) programming to allocate different levels of power to the multiplexed users and the subchannels. To reduce the complexity of the decoding at receivers, the number of users allocated on the same sub-band was limited to two. The NOMA system equipped with the proposed resource allocation scheme can achieve higher sum rate and higher energy efficiency than the traditional OFDMA system. In OFDMA heterogeneous networks, users in the same small cell are not allowed to share the same frequency band in order to avoid the inter-cell interference \cite{haijunIEEE14}.

In the NOMA heterogeneous networks, NOMA can be implemented in each small cell, which means that the users in the same small cell are allowed to share the same sub frequency band. Different from NOMA single cell system, users in the same small cell can share the same sub-band with the users in macrocell. The interference signal from the users in the small cell can be detected  with the superposition coding technique and removed by the strong users in the macrocell via SIC at the receivers.
\subsection{NOMA Heterogeneous Network Model}
In this system, we assume each small cell occupies one sub frequency band. It means that all the users in the same small cell can be multiplexed on the same sub frequency band. The MUEs can be co-multiplexed on the same sub frequency band with SUEs. Each BS is responsible for frequency bands assignment (user scheduling), access procedures and power allocation.

\begin{figure}[h]
        \centering
        \includegraphics*[width=13cm]{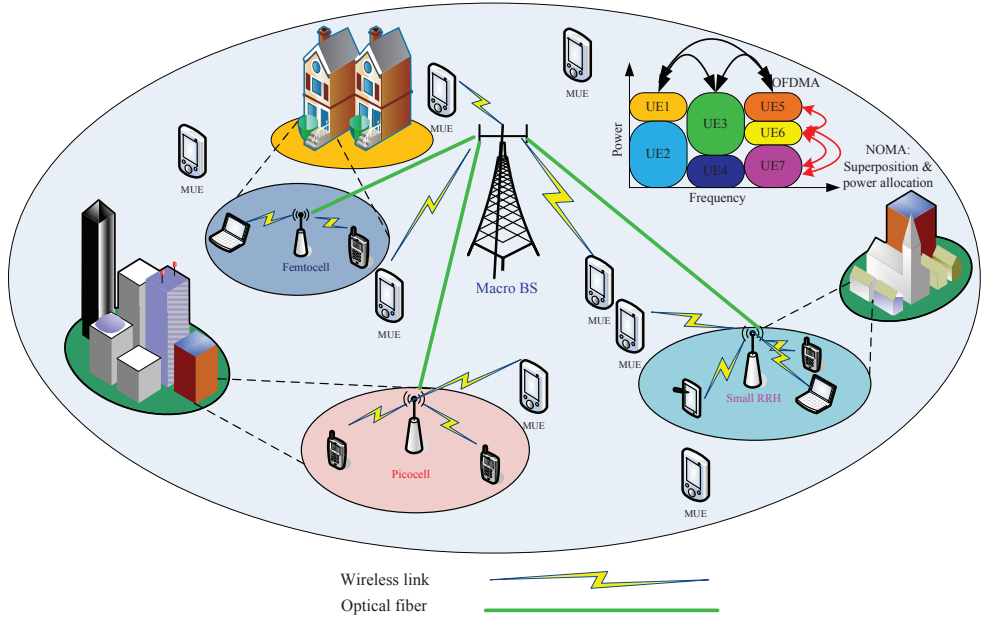}
        \caption{NOMA based Heterogeneous Networks.}
        \label{fig:5}
\end{figure}
In the following, we use sub-band to replace sub frequency band for the sake of simplicity. In NOMA heterogeneous networks, each UE can receive different signals from the macro BS and a small BS simultaneously. Furthermore, each BS transmits signals to UEs by using superposition coding. As shown in Fig. 1, we consider a downlink NOMA heterogeneous network where a primary macrocell hosts a macro base station (MBS) with $K$ overlaid small cells. Let $M$ denote the total number of active MUEs in the macrocell. Each small cell hosts one small BS (SBS) and $F$ SUEs that are randomly distributed within the small cell. The MBS transmits signals to $M$ MUEs through these $K$ sub-bands. The total bandwidth of the system is denoted as $B$, which is divided into $K$ sub-bands and each sub-band occupies bandwidth of $B/K$. The transmit power for the entire system is $P_s$. Because the transmit power for MBS should be higher than SBS, we assume that both the MBS and all the SBSs have transmit power $P_s/2$ and each SBS has transmit power $P_s/(2K)$. A block fading channel is considered in the system model, where the channel fading of each subcarrier is assumed to be the same within a sub-band, but it varies independently across different sub-bands. In this system, the MUEs can share the same frequency band with the small cells. The cross-tier interference caused by the primary macrocell will be effectively mitigated by applying the NOMA technique. Note that interference between small cells is neglected because different small cells will occupy different sub-bands and the sub-bands are independent with each other. The small cell is allocated lower power than the macrocell and the signal within the small cell can suffer severe wall penetration loss \cite{haijunIEEE14}. Therefore, the co-tier inter-small cell interference is negligible when compared with cross-tier interference \cite{haijunIEEE14}. We assume that MBS and SBS are connected by the wired links, thus the backhaul will not be considered in this article.

The resource optimization problem is formulated as an  energy efficiency maximization problem for the downlink NOMA heterogeneous networks. Our goal is to maximize the total system sum rate with a unit power through resource scheduling. In other words, in order to guarantee amount of system data rate, we aim to reduce the power consumption as much as possible. In our study, energy efficiency is defined as the ratio of the system sum rate and total power consumption. The total power consumption includes both circuitry-consumed power of the mobile devices and transmission power. Therefore, we use bits per Joule to measure the system energy efficiency performance. The energy-efficient resource optimization problem can be formulated as the maximization of the energy efficiency. In this problem, user scheduling and power allocation are coupled each other. In order to solve this problem efficiently, the resource allocation for energy efficiency maximization problem can be divided into user scheduling and power allocation sub-problems. In the following, we will discuss the resource optimization for NOMA heterogeneous networks based on two cases: perfect CSI and imperfect CSI.

\section{Energy Efficient Resource Optimization for NOMA Heterogeneous Networks with Perfect CSI}

We first consider the BSs have knowledge of channel state information. For user scheduling, matching theory can be utilized to match different UEs with different sub-bands. In the proposed energy efficient subchannel assignment framework \cite{FangIEEETrans16} for a downlink NOMA single cell network, the utility function of each user is defined as the summation of each subchannel energy efficiency which can be described as the ratio of the subchannel data rate and power consumption. The power consumption for each subchannel includes circuit power and transmission power on this subchannel. In the NOMA heterogeneous networks, we define our user scheduling optimization sub-problem as the summation of small cells energy efficiency. For user scheduling scheme design, we first assume equal power allocation on each sub-band. This user scheduling optimization is a non-convex mixed integer programming problem. The global optimal user scheduling scheme can be obtained by an exhaustive search. However, the complexity of the exhaustive search is extremely high and this approach is infeasible in practice. Assume we have $M$ MUEs and $N$ small cells.  The scheduler needs to search $\frac{{(M)!}}{{{2^N}}}$ combinations. Therefore, a suboptimal user scheduling scheme can be designed for this system by using a matching theory. In the sub-optimal scheme, the complexity of the worst case is $O({N^2M})$. Taking natural logarithm of the complexity, the logarithm complexity is $O(\ln M)$. Since $O(\ln M)<O(M\ln M)$. The complexity of the sub-optimal scheme is much less than the exhaustive searching. In the proposed user scheduling scheme, we consider all the UEs as a user set, and consider $K$ sub-bands as a sub-band set. The user scheduling process can be described as the two-sided dynamic matching between the user set and the sub-band set. For each sub-band, we set a maximum number of users that can be multiplexed on one sub-band in order to reduce the decoding complexity at the receivers. Therefore, the energy efficiency of the overall system can be formulated as the maximization of energy efficiency under the constraints of maximal transmission power, the minimum user data rate as well as the maximum number of users can be allocated on one sub-band. Fig. 2 describes the resource optimization procedure including user scheduling and power allocation.
\begin{figure}[h]
        \centering
        \includegraphics*[width=13cm]{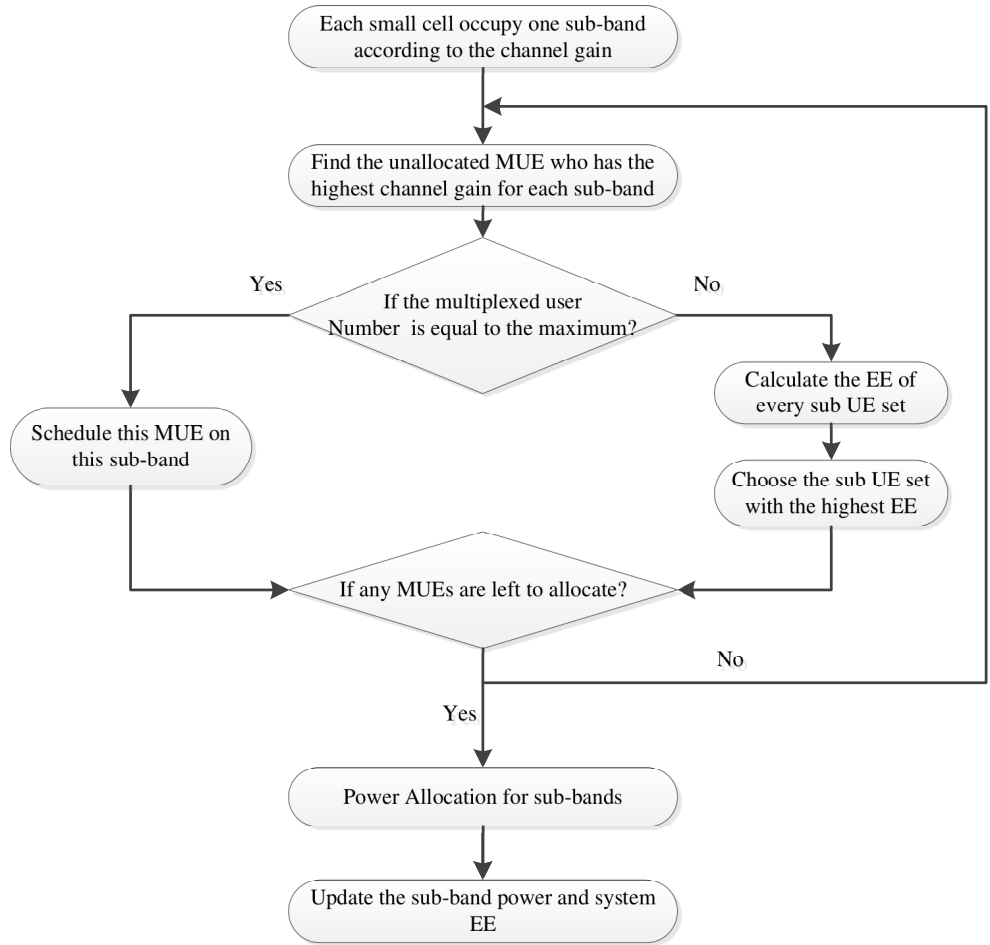}
        \caption{Energy efficient resource optimization procedure.}
        \label{fig:2}
\end{figure}

\subsection{User Scheduling}

In user scheduling, we consider the SUEs scheduling and the MUEs scheduling separately. Since NOMA is implemented in each small cell, all the SUEs in one small cell can only multiplex on one sub-band. Then SUEs scheduling can also be called small cells scheduling. In the small cells scheduling, since the scheduler has the perfect CSI, each sub-band has preferred sequence of small cells to assign. This preferred sequence is generated according to the channel gains between SBS and SUEs within each small cell. Assume, for the time being, the equal power allocation among SUEs in the small cell, the energy efficiency of the small cell on each sub-band can be calculated by the scheduler. The scheduler chooses the sub-band with the highest channel gain and schedule the corresponding small cell on this sub-band. After all the small cells have occupied sub-bands, the scheduler schedules the MUEs on these occupied sub-bands, and the MUEs scheduling depends on the energy efficiency that the MUE can provide for the sub-bands. Therefore, the SUEs and MUEs co-multiplex on these sub-bands.
In the MUEs scheduling scheme, each sub-band has a preferred allocation list of MUEs in a decreasing order of the channel
gains between MBS and MUEs. For each sub-band, the scheduler finds the highest channel gain and schedule its corresponding
MUE onto this sub-band. According to the condition of the maximum number of sub-band users, the MUE allocates on the
sub-band directly when the number of the multiplexed users on this sub-band is less than the maximum number. However, when
the number of the multiplexed users on this sub-band equals the maximum, the scheduler compares the energy efficiency
of users and assign the sub-band to the users who can provide the highest energy efficiency.
In the energy efficiency calculation, we assume the power of MUEs equals the power of small cell on this sub-band and each SUE has the equal power. The MUEs matching process terminates when all the UEs have been multiplexed on the sub-bands. The optimal sub-band matching can only be obtained by searching all matching combinations of users and sub-bands. This procedure comes with extremely high complexity and is not practical. The above proposed sub-optimal user scheduling has lower complexity than the optimal solution.
\subsection{Power Allocation}
In user scheduling, we have assumed the equal power allocation across sub-bands and equal power for small cells and MUEs on the same sub-band. In order to further improve the system energy efficiency, sophisticated power allocation schemes can be proposed to replace the equal power allocation scheme. In the NOMA heterogeneous networks, the energy efficiency optimization problem is non-convex. To solve this problem, DC programming approach and time-sharing method can be used to transform the non-convex combinatorial optimization problem into a convex optimization problem. In this article, we exploit fractional transmit power allocation (FTPA) to further allocate powers among users and across sub-bands \cite{FangIEEETrans16}. Due to its low computational complexity, FTPA has been widely adopted in the OFDMA and NOMA systems.

\begin{figure}[h]
        \centering
        \includegraphics*[width=13cm]{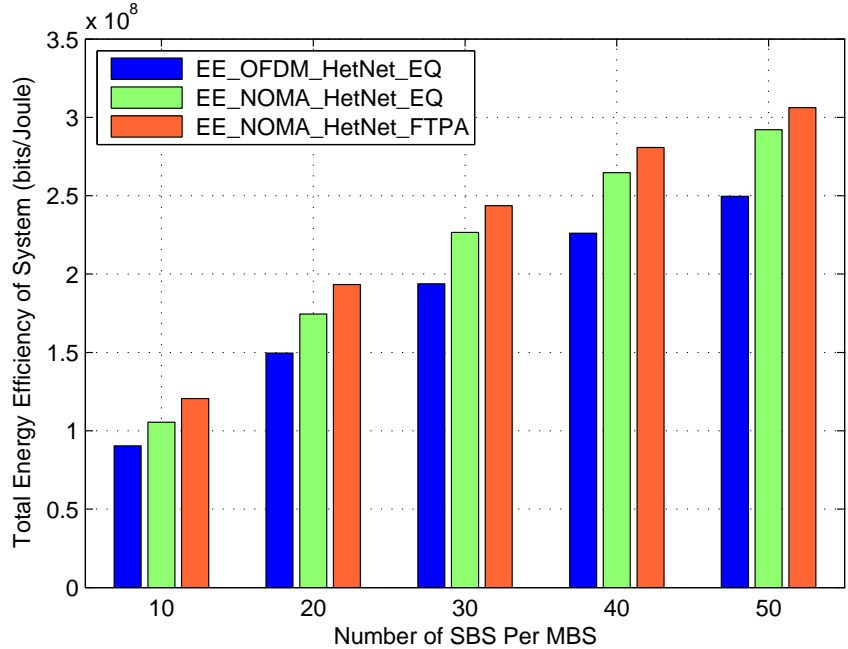}
        \caption{Energy efficiency vs. number of  small cells.}
        \label{fig:3}
\end{figure}

Figure 3 shows the performance of the system energy efficiency versus the number of the small cells. The bandwidth is limited to 10 MHz, and we set the peak power of the entire system to be 20 W and circuit power to be 0.1 W on each sub-band. Figure 3 indicates that the total energy efficiency increases when the number of the small cells increases. From Fig. 3, we can observe that NOMA heterogeneous network (NOMA HetNet) outperforms the OFDM heterogeneous network (OFDM HetNet) because OFDM HetNet cannot fully utilize the spectrum resources.
 \begin{figure}[h]
        \centering
        \includegraphics*[width=13cm]{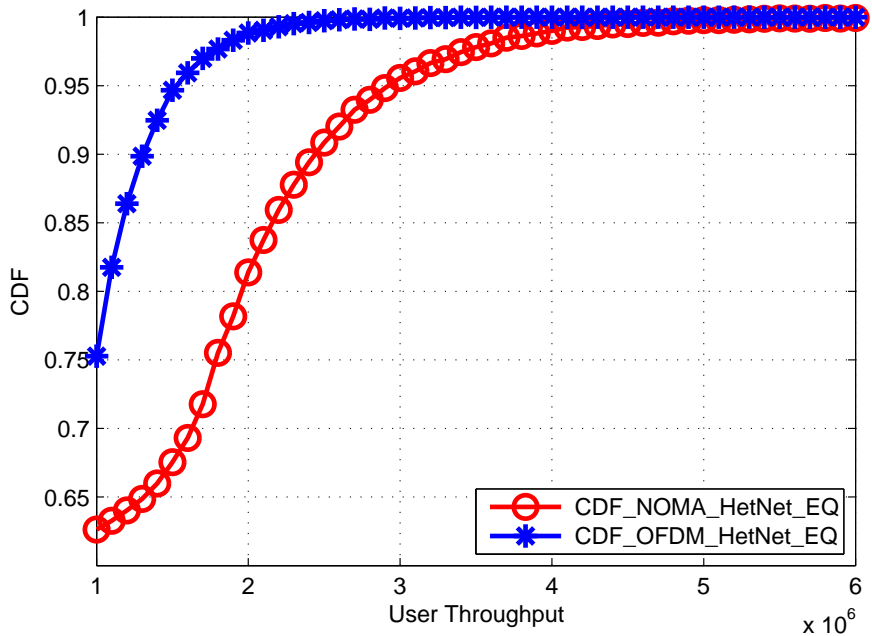}
        \caption{CDF of data rate for NOMA HetNet and OFDMA HetNet.}
        \label{fig:4}
\end{figure}

Fig. 4 shows the cumulative distribution function (CDF) of user average data rate comparison between NOMA HetNet and OFDM HetNet with equal power allocation scheme. In Fig. 4, the CDF of user average data rate in NOMA HetNet is always below that in OFDM HetNet. This means that, given a target data rate, the number of users whose average data rate is larger than this target data rate in NOMA HetNet is less than that in NOMA HetNet system. This is because NOMA technique has higher spectrum efficiency than the OFDMA scheme.

\section{Energy Efficient Resource Optimization for NOMA Heterogeneous Network with Imperfect CSI}
In this subsection, we study the energy efficient resource allocation for the downlink NOMA heterogeneous network with imperfect CSI. In practice, the perfect CSI at the transmitter is difficult to achieve due to channel estimation errors, feedback and quantization errors. To maximize the energy efficiency, we can formulate the energy efficient resource allocation as a probabilistic mixed non-convex optimization problem, under the constraints of outage probability limitation, the maximum transmitted power and the number of multiplexed users on one sub-band.

The resource allocation scheme is optimized based on the imperfect CSI. To solve this problem, we decouple the problem into user scheduling and power allocation sub-problems separately to maximize the system energy efficiency.
We assume that the BSs have the estimated value of the CSI. In this situation, the user data rate may not meet the minimum data requirement determined by QoS. Therefore, an outage probability requirement is considered for the resource scheduling to maximize the system energy efficiency.
The energy efficient optimization problem is formulated as a probabilistic mixed non-convex optimization problem. In order to efficiently solve the optimization problem, we can first transform the probabilistic mixed non-convex optimization problem to a non-probabilistic optimization problem. The outage probability constraint can be transformed to minimum power constraints for each UEs sharing the same sub-band. In this transformed problem, we can treat user scheduling and power allocation separately. The user scheduling starts with assigning equal power allocation on sub-bands. The optimal solution to the user scheduling sub-problem is challenging to obtain in practice as it requires to search all the possible combinations. A sub-optimal user scheduling algorithm can be obtained by using the estimated CSI. In order to cancel the interference from the other small cells, we assign each small cell a sub-band. MUEs  multiplex on these sub-bands, and hence the small cell users  not interfere with each other. The MUE with high channel gain is chosen to multiplex on its corresponding sub-band. For the each sub-band, we set the maximum number of multiplexed users to reduce the complexity of decoding at the receivers. The user scheduling  terminates if there is no MUE left to be allocated. During the user scheduling, the power allocation for the multiplexed users on one sub-band can be determined by FTPA. The complexity of the proposed sub-optimal algorithm is less than the optimal solution that can only be obtained by the exhaustive search.
  Given the user scheduling scheme, the underlying optimization problem can be shown to be convex function respect to the power variable under the constraint of the maximum power of the system. Therefore, a unique optimal solution can be found by a gradient assisted binary search algorithm. Therefore, the new power allocation scheme can replace the equal power allocation to further improve the system energy efficiency.

\begin{figure}[h]
        \centering
        \includegraphics*[width=13cm]{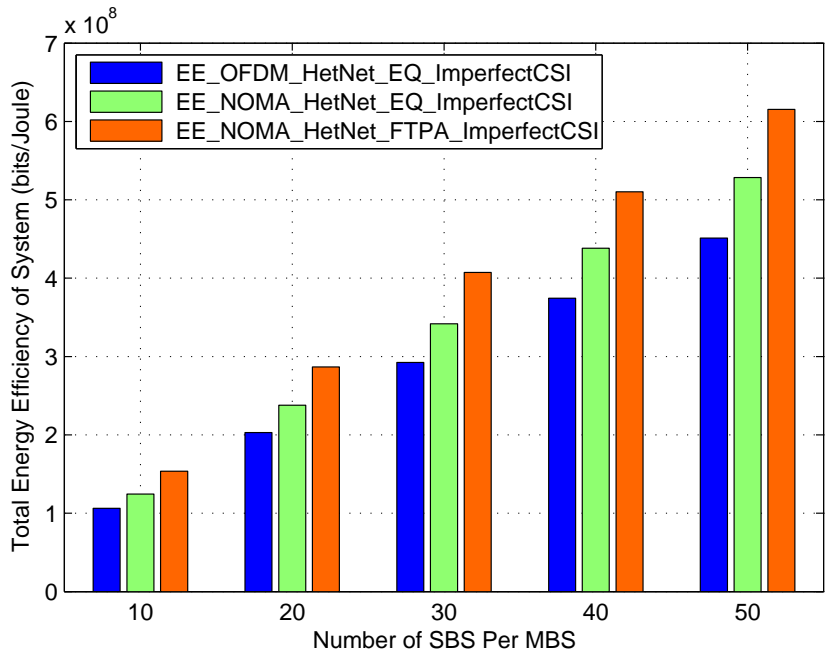}
        \caption{Energy efficiency of the system vs. the number of the small cell.}
        \label{fig:5}
\end{figure}

Figure 5 shows the energy efficiency versus the number of the small cells with the channel gain estimation error variance 0.05. As shown, the energy efficiency increases when the number of the small cells grows. From Fig. 5, the NOMA HetNet outperforms OFDM HetNet in terms of energy efficiency, and the energy efficiency of NOMA HetNet with FTPA is higher than that of scheme with equal power allocation.
\begin{figure}[h]
        \centering
        \includegraphics*[width=13cm]{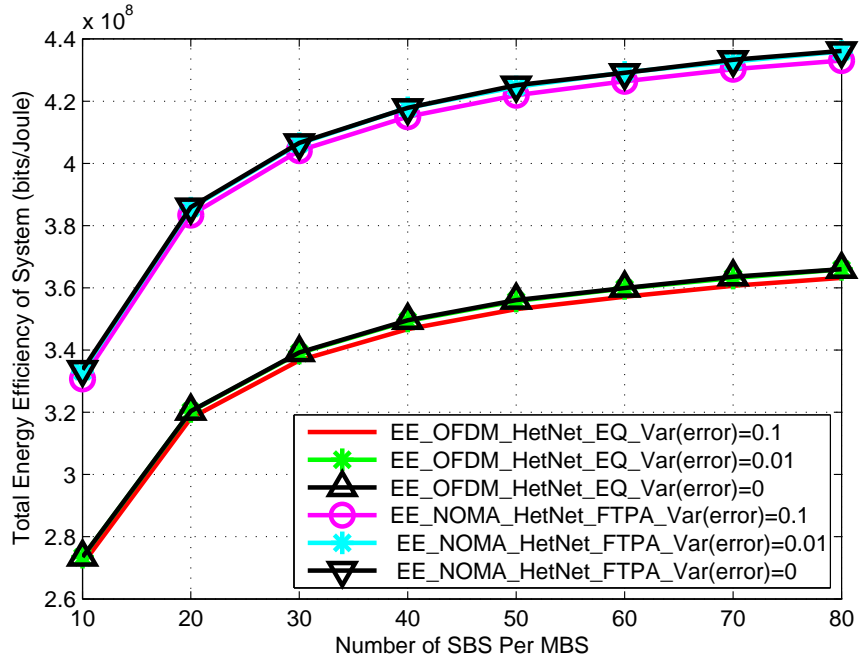}
        \caption{Energy efficiency of the system vs. the number of the small cell.}
        \label{fig:6}
\end{figure}

Figure 6 shows the energy efficiency of the NOMA HetNet performance with different estimation error variances. In both OFDM HetNet and NOMA HetNet schemes, the system energy efficiency deteriorates when the error variances increases. The NOMA HetNet always outperforms OFDM HetNet. For example, when the number of small cells is 20, the energy efficiency of NOMA HetNet with FTPA power allocation scheme is 27\% more than that of the OFDM HetNet with the equal power (EQ) allocation scheme under the imperfect channel CSI.

\section{Open Issues, and Challenges}
Resource allocation is one of the methods to improve the energy efficiency of 5G NOMA networks. Small cell in HetNet can also be energy efficient because low power is typically required due to the short distance between transmitter and receiver. However, there are still many challenges and open issues in energy-efficient 5G NOMA networks. 1) NOMA with MIMO: Not only user scheduling and power allocation, but also beamforming can be considered in the energy efficient resource allocation. Furthermore, different configuration of antennas for macrocell and small cell in HetNet makes the problem more complex. Finding effective ways to combine NOMA and massive MIMO is a challenging but important topic in future research works. 2) NOMA with energy harvesting: To save more energy, 5G NOMA HetNet is expected to harvest energy from solar, wind, thermoelectric, and so on. One open problem is  how to manage those renewable energy source because they are intermittent over time and space. 3) NOMA with game theory: User pairing is key research direction in energy-efficient 5G NOMA networks. Matching game theory can be an effective tools to distributed model and optimize the user pairing of energy-efficient 5G NOMA networks. 4) NOMA with open/closed access small cells: There are three access modes in  small cell networks, open access, closed access, and hybrid access. Different access mode of NOMA small cells requires different energy saving methods.

\section{Conclusion}

We introduced the concept of NOMA heterogeneous network and examined the problem of energy efficient user scheduling and power allocation in 5G NOMA heterogeneous network by considering both perfect CSI and imperfect CSI. We solved the resource optimization problem with the help of convex optimization. Based on the obtained results, we proposed a two-step user scheduling and power optimization scheme. The effectiveness of the proposed schemes was compared to the existing scheme and verified by simulations in terms of energy efficiency. It is envisioned that a joint user scheduling and power allocation approach can further enhance the energy efficiency of the overall system.

\end{document}